\documentclass[onecollarge]{svjour2}       % onecolumn "king-size"
\smartqed  % flush right qed marks, e.g. at end of proof
\usepackage{graphicx}
\begin{document}

\title{Fiber bundle model under heterogeneous loading}
\subtitle{}


\author{Subhadeep Roy \and Sanchari Goswami}


\institute{Subhadeep Roy \at
              Earthquake Research Institute, University of Tokyo, 1-1-1 Yayoi, Bunkyo, 113-0032 Tokyo, Japan. \\
              \email{sroy@eri.u-tokyo.ac.jp}        
           \and
           Sanchari Goswami \at
             Vidyasagar College, 39 Sankar Ghosh Lane, kolkata- 700006, India \\
              \email{sg.phys.caluniv@gmail.com}
}

\date{Received: date / Accepted: date}

\maketitle

%%%%%%%%%%%%%%%%%%%%%%%%%%%%%%%%%%%%%%%%%%%%%%%%%%%%%%

\begin{abstract}
The present work deals with the behavior of fiber bundle model under heterogeneous loading condition. The model is explored both in the mean-field limit as well as with local stress concentration. In the mean field limit, the failure abruptness decreases with increasing order $k$ of heterogeneous loading. In this limit, a brittle to quasi-brittle transition is observed at a particular strength of disorder which changes with $k$. On the other hand, the model is hardly affected by such heterogeneity in the limit where local stress concentration plays a crucial role. The continuous limit of the heterogeneous loading is also studied and discussed in this paper. Some of the important results related to fiber bundle model are reviewed and their responses to our new scheme of heterogeneous loading  are studied in details. Our findings are universal with respect to the nature of the threshold distribution adopted to assign strength to an individual fiber.

%\pacs{05.70.Fh}{Phase transitions: general studies}
%\pacs{05.40.Ca}{Noise}
%\pacs{62.25.MN}{Fracture/brittleness.
\end{abstract}

%%%%%%%%%%%%%%%%%%%%%%%%%%%%%%%%%%%%%%%%%%%%%%%%%%%%%%%

\section{Introduction}
Fracture in materials is a complex phenomenon which involves very large length and time scales. Fracture in materials are mainly guided by either extreme events or collective behavior of the defects present in the material. These modes of failure depends on many parameters like temperature \cite{Wonga}, pressure \cite{Wonga}, lattice defects \cite{Brede,Gumbsch}, porosity \cite{Li}, strain rate \cite{Khantha1,Khantha2} etc. In last few years, there are many attempts in statistical mechanics to include such effects and understand the failure process from numerical point of view (specifically in random spring network \cite{Curtin1,Curtin2} and random resistor network \cite{Nukula,Khang,Shekhawat,Moreira} model). One of such model is fiber bundle model (FBM) \cite{Chak2,Fiber1,Fiber2} that has been proven to show many aspects of failure process in previous years. 

Fiber bundle model, after its introduction in 1926 by Pierce \cite{Pierce}, has been explored extensively. The model is a classic example of a disordered system out of equilibrium, mainly guided through the weakest link of a chain concept \cite{Tanaka,McClintock,Ray}. Previous studies in the mean-field limit of the model have shown universal behavior like a scale-free avalanche with an unique exponent $-5/2$ \cite{Hemmer}. On the other hand, with local stress concentration, a logarithmic decrease in bundle strength claimed with increasing system sizes, both analytically \cite{lls1} and numerically \cite{lls2}. So far the model is mainly observed under homogeneous loading condition. In this paper we propose a different algorithm for FBM where the loading is heterogeneous.       

In this paper the mean field limit of the model is explored first, followed by the study with local stress concentration. In mean-field limit, the fluctuation in stress redistribution is ignored and therefore the model is considered to be operated with a global load sharing scheme \cite{Daniels}. The other spectrum is obviously the local load sharing scheme \cite{Phoenix1,Phoenix2,Phoenix3,phoenix09,Harlow1,Harlow2,Harlow3} where the stress redistribution is very localized and hence heterogeneous. Instead of the usual local load sharing scheme, the load redistribution over a range can also be studied which is done recently in ref. \cite{soumya}. In case of heterogeneous loading, the stress profiles on the fibers are different and they occur according to the order of heterogeneity in loading process. 

There are some previous works that deals with the origin of non-uniform stress, comes into the play due to flaws in real structures . A series of studies are performed on semielliptical surface cracks \cite{Raju,Newman,Vainshtok,Wang,Fett} in a flat plane because of its application in idealizing the flaws in real systems. The works by Raju and Newman \cite{Raju,Newman} are most acceptable in such cases. In ref.\cite{Vainshtok}, the average error in the stress intensity factor, due to flaws in real structures, was evaluated using the concept of energy release rate. Similar to above discussed studies, in the present paper also a non-uniform local stress arises within the model even when a particular stress is externally applied on it. The fiber bundle is being studied under non-uniform tensile stress earlier \cite{phoenix73,phoenix75,phoenix74,daniels89}. In those studies mostly the fibers are considered to be composed of several sub-bundles (random fiber slack effect \cite{phoenix79}), associating random variables withing the loading process and rupture events of any two fibers from a same sub-bundle is allowed to be probabilistically dependent.           

In the next section we have described the FBM in details. Sec.3 and Sec.4 contains the analytical and numerical findings with such heterogeneous loading scheme. One of the main attempt is to understand how the mode of failure is affected by the order $k$ of heterogeneity in loading process. In the mean field limit we have explored the brittle to quasi-brittle transition point, the critical point separating the abrupt from the non abrupt failure, as a function of $k$. In the local load sharing (LLS) limit, we have discussed the system size effect of failure strength as well as the correlation in rupture events. Above properties in the LLS limit has been already discussed in Ref. \cite{lls2} and \cite{soumya}. Here we have revisited the studies to understand the effect of heterogeneous loading. The final part of the numerical result is contributed to the study of the continuous limit of such heterogeneity. Finally, in Sec.5 and Sec.6 we have given a brief discussion on our findings and supporting evidence for the universality of results respectively.          

%%%%%%%%%%%%%%%%%%%%%%%%%%%%%%%%%%%%%%%%%%%%%%%%%%%%%%%

\section{Description of the Model}
\begin{figure}[ht]
\centering
\includegraphics[width=7cm, keepaspectratio]{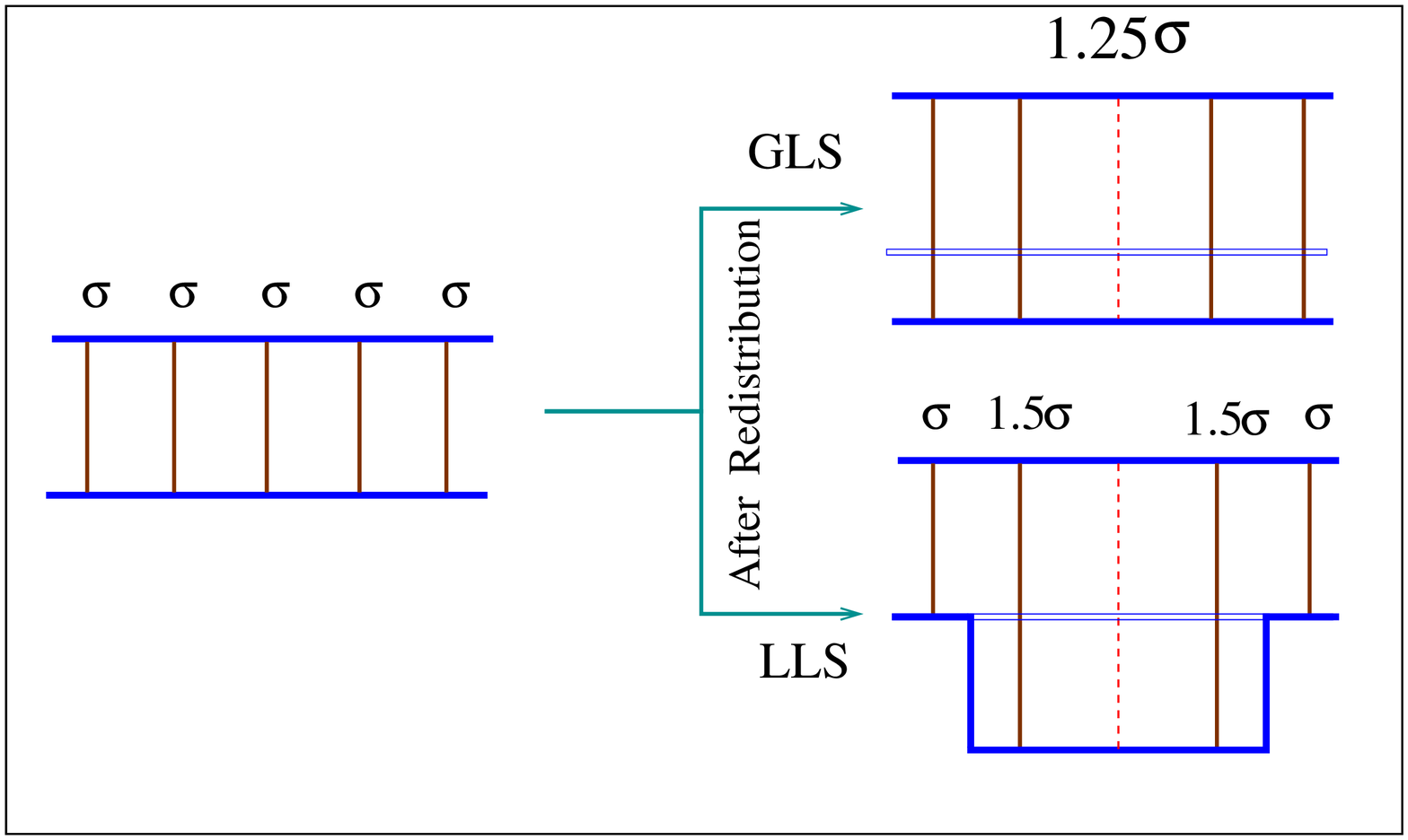} \ \ \ \ \ \ \  \vspace{0.3cm} \includegraphics[width=4.5cm, keepaspectratio]{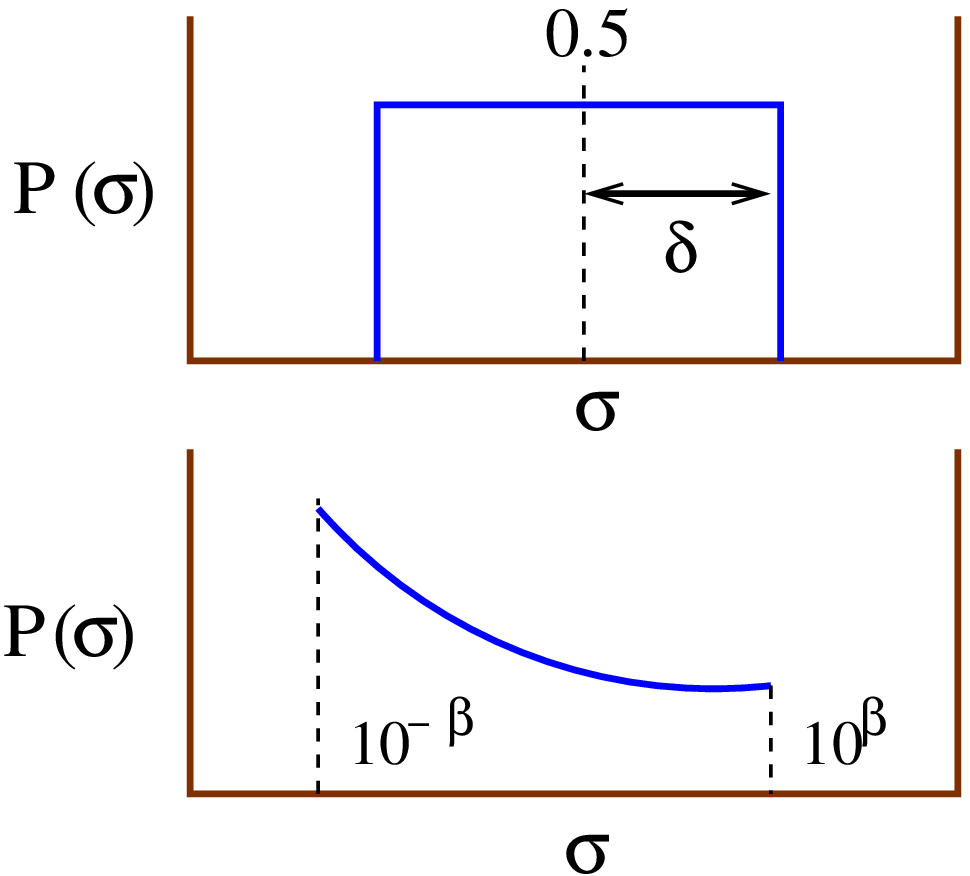}
\caption{Left: Fiber bundle model with global (GLS) and local (LLS) sharing scheme. The GLS scheme is also the mean field limit. In LLS limit a high stress concentration is observed near a broken fiber. Right: Examples for the threshold strength distribution $P(\sigma)$. For numerical results we use this uniform distribution of half-width $\delta$ and mean at 0.5. The universality of the results are confirmed from another threshold distribution: a power law with slope -1 from $10^{-\beta}$ to $10^{\beta}$. $\delta$ and $\beta$ measures the strength of disorder.}
\label{fig:fbmThres}
\end{figure}

The basic Fiber Bundle Model (FBM) (see Fig.\ref{fig:fbmThres}) is a simple yet useful model to study fracture-failure phenomena. It was first introduced by Pierce \cite{Pierce}. The model consists of vertical fibers ($L$ fibers) between two horizontal bars (see Fig.\ref{fig:fbmThres}). The bars are pulled apart by a force $F$, creating an external stress per fiber $\sigma$ (=$F/L$). Disorder is introduced in the model as the fluctuation in strength of individual fibers. The threshold strength values (\{$\sigma$\} values) are chosen randomly from a certain distribution. Fig.\ref{fig:fbmThres} shows two such distributions : uniform and power law.
In conventional fiber bundle model, initially a constant stress $\sigma$ (described above) is applied on all the fibers. In the present work we have mainly studied model when a non uniform stress is applied creating a heterogeneity loading condition. Due to such heterogeneous stress increment scheme, the local stress per fiber profile is different from that of the external stress per fiber $\sigma$. The algorithm for the evolution of the model is described below: 

\begin{figure}[ht]
\centering
\includegraphics[width=9cm, keepaspectratio]{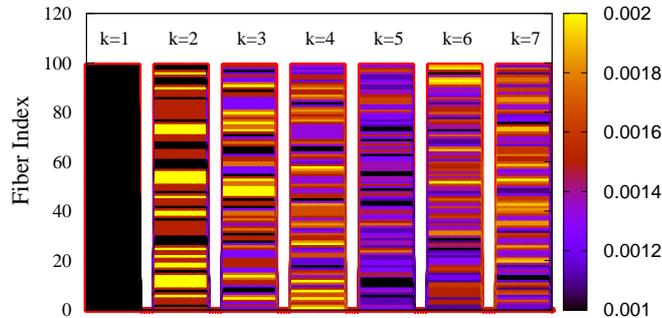}
\caption{Local stress profile of the fibers when an external stress 0.001 is applied on it. $k$ is the order of heterogeneity in the loading process. The maximum amplification factor is chosen to be 2.}
\label{fig:Phase_Diagram0}
\end{figure}

\begin{enumerate}
\item Let's assume that there is an heterogeneity of order $k$ in the loading process. For an externally applied stress $\sigma$, the local stress $\sigma(i)$ of individual fibers may assume $k$ different equispaced values with probability $1/k$. At the same time each fiber is accompanied by local parameter $\alpha(i)$ which acts as an amplification factor over the applied stress. For numerical study we have restricted the $\alpha(i)$ values within 1 and 2. Example: for $k=3$ the possible $\alpha(i)$ values will be 1.0, 1.5 and 2.0; for $k=4$ such values will be 1.0, 1.33, 1.66 and 2.0. For the sake of stress conservation we have adopted the following rule to express the local stress as a function of external stress:    
\begin{equation}\label{eq0}
\sigma(i)=[\sigma\alpha(i)]-\displaystyle\frac{\displaystyle\sum_{i}[\sigma\alpha(i)]-\displaystyle\sum_{i}\sigma}{N_u}
\end{equation}    
where $\sigma$ is the externally applied stress and $N_u$ is number of intact fibers. $i$ runs over all intact fibers. Fig.\ref{fig:Phase_Diagram0} shows the local stress profile for a bundle with $10^2$ fibers while a stress $\sigma=0.001$ is applied externally on it. $k$ is the heterogeneity in loading process. $k=1$ leads to the conventional model where all fibers experiences same stress $\sigma$. As we increase the $k$, different stress profiles will arise between 0.001 and 0.002, shown through different colors in the figure. \\
\item A fiber breaks irreversibly if the local stress $\sigma(i)$ matches with its threshold value. As an effect of the heterogeneous loading, a fiber with high $\alpha(i)$ value might break even when the applied stress $\sigma$ is less than its threshold strength. \\ 
\item After each rupture event, the stress of a broken fiber is redistributed within the rest of the bundle. Fig.\ref{fig:fbmThres} shows two such stress redistribution schemes where either the stress is redistributed among all surviving fibers (GLS, global load sharing \cite{Pierce,Daniels}) or in the local neighborhood only (LLS, local load sharing \cite{Phoenix1,Phoenix2,Phoenix3,Harlow1,Harlow2,Harlow3}). 

Fig.\ref{fig:fbmThres} illustrates the redistribution rule for a bundle with $L=5$. Let's assume that a stress $\sigma$ is applied on the fibers at a certain point. Now if a breaks then the redistributed stress will be as follows: 
\begin{itemize}
\item GLS $\Rightarrow$ All fibers will carry a stress $\sigma+(\sigma/4)=1.25\sigma$, as the extra stress will be carried by 4 unbroken fibers.  
\item LLS $\Rightarrow$ In this case the extra stress will be redistributed among the nearest neighbors (2 fibers) and hence they will carry a stress $\sigma+(\sigma/2)=1.5\sigma$. Stress applied on other fibers will be $\sigma$.
\end{itemize}
A recent study \cite{soumya} on the model discusses the scheme where the stress is redistributed over a range $R$. Such nature of redistribution actually depends on the stiffness of the horizontal bars; higher stiffness leads to the mean field limit while for very low stiffness the stress will be redistributed locally. \\ 
\item After such redistribution the local stress profile of certain fibers, that carries the extra load of the broken one, get modified and there may be further breaking without increasing the external stress. This procedure stops when the redistributed stress does not reach the next threshold value. \\
\item During above redistribution the bundle might fail catastrophically through avalanches or it comes to a stable state after a few ruptures. In the later case, the external stress is raised to create the next rupture and the chain of redistribution starts again. The stress increment also follows Eq.\ref{eq0}. If $\Delta\sigma$ is the external stress increment, then the local increment in stress will be given by:  
\begin{equation}\label{eq00}
\Delta\sigma(i)=[\Delta\sigma\alpha(i)]-\displaystyle\frac{\displaystyle\sum_{i}[\Delta\sigma\alpha(i)]-\displaystyle\sum_{i}\Delta\sigma}{N_u}
\end{equation}    
We keep increasing this external stress until all fibers break suggesting the global failure.    
\end{enumerate}

We have studied the mean field limit of the model both analytically and numerically with varying order of heterogeneity. 
The model is also observed numerically with varying order of heterogeneity together with disorder and stress release range.            

%%%%%%%%%%%%%%%%%%%%%%%%%%%%%%%%%%%%%%%%%%%%%%%%%%%%%%%

\section{Analytical Approach}
The model is studied analytically under heterogeneous stress increment (or loading) but homogeneous (global) stress 
redistribution (mean field limit).
We have chosen a heterogeneity of 
order $k$ that will lead to $k$ different local stress values. 
Due to 
heterogeneity in local stress, the fibers may be grouped into $k$ types of fibers with local stress values 
$\sigma_1$, $\sigma_2$, $\cdots$, $\sigma_k$ with equal probability $1/k$.  
At a certain applied stress $\sigma$, if $n_{b1}$, $n_{b2}$, $\cdots$, $n_{bk}$ 
fraction of fibers of type $1$, $2$, $\cdots$ $k$ are broken, then the local stress profile after 
redistribution will be given by
\begin{equation}\label{eq1}
\sigma_{rj}=\sigma_j+\displaystyle\frac{\displaystyle\sum_{i=1}^{k}n_{bi}\sigma_i}{\left[1-\displaystyle\sum_{i=1}^{k}n_{bi}\right]}
\end{equation}
where $\sigma_j$ and $\sigma_{rj}$ are the local stress on the fibers of type $j$, before and after redistribution. 
Now, fraction of unbroken fibers of type $j$ can be calculated by integrating the threshold distribution from 
$0$ to $\sigma_{rj}$. In our analytical calculation we have considered uniform distribution, given by 
$p(\sigma)=1/2\delta$ ($\delta$ is the half width of the distribution). 
\begin{equation}\label{eq2}
n_{bj}=\frac{1}{k}\displaystyle\int^{\sigma_{rj}}_{a}p(\sigma)d\sigma =\frac{1}{2k\delta}(\sigma_{rj}-a)
\end{equation}
where $a$ corresponds to the minimum of the threshold distribution. Inserting the value of $\sigma_{rj}$  we get
\begin{equation}\label{eq3}
n_{bj}=\left(\frac{1}{2k\delta}\right)\left[\displaystyle\frac{\sigma_j-\sigma_j\displaystyle\sum_{i=1}^{k}n_{bi}+\displaystyle\sum_{i=1}^{k}n_{bi}\sigma_i-a+a\displaystyle\sum_{i=1}^{k}n_{bi}}{1-\displaystyle\sum_{i=1}^{k}n_{bi}}\right]
\end{equation}
Separating the the contribution due to $j^{th}$ part from rest of the mixture, we get 
\begin{equation}\label{eq4}
n_{bj}=\left(\frac{1}{2k\delta}\right)\displaystyle\frac{\Bigg[\sigma_j-a+a\displaystyle\sum_{i \ne j}^{k}n_{bi}-an_{bj}- \nonumber \sigma_j\displaystyle\sum_{i \ne j}^{k}n_{bi}+\displaystyle\sum_{i \ne j}^{k}n_{bi}\sigma_i\Bigg]}{\Bigg[1-\displaystyle\sum_{i \ne j}^{k}n_{bi}-n_{bj}\Bigg]}
\end{equation}
Simplifying Eq. \ref{eq4} we get a quadratic equation of $n_{bj}$ 
\begin{equation}\label{eq5}
n_{bj}^2-n_{bj}\Bigg[1-\displaystyle\sum_{i \ne j}^{k}n_{bi}-\displaystyle\frac{a}{2k\delta}\Bigg]+\displaystyle\frac{a}{2k\delta}\Bigg[\sigma_j-a+a\displaystyle\sum_{i \ne j}^{k}n_{bi}-\sigma_j\displaystyle\sum_{i \ne j}^{k}n_{bi}+\displaystyle\sum_{i \ne j}^{k}n_{bi}\sigma_i\Bigg]=0
\end{equation}
Solution of above equation will be 
\begin{equation}
\hspace{-1.3cm}n_{bj}=\displaystyle\frac{1}{2}\left[\left(1-\displaystyle\sum_{i \ne j}^{k}n_{bi}-\displaystyle\frac{a}{2k\delta}\right)\pm\sqrt{\left(1-\displaystyle\sum_{i \ne j}^{k}n_{bi}-\displaystyle\frac{a}{2k\delta}\right)^2-\displaystyle\frac{a}{2k\delta}\left(\sigma_j-a+a\displaystyle\sum_{i \ne j}^{k}n_{bi}-\sigma_j\displaystyle\sum_{i \ne j}^{k}n_{bi}+\displaystyle\sum_{i \ne j}^{k}n_{bi}\sigma_i\right)}\right]
\end{equation}
Above mentioned solution will reduce to the following form at critical point :  
\begin{equation}\label{eq6}
n_{bj}^c=\frac{1}{2}\left[1-\displaystyle\sum_{i \ne j}^{k}n_{bi}-\displaystyle\frac{a}{2k\delta}\right]
\end{equation}
Then, fraction unbroken at this point will be
\begin{equation}\label{eq7}
n_u^c=1-\displaystyle\sum_{j=1}^{k}n_{bj}^c=\displaystyle\frac{1}{k+1}\left(1+\displaystyle\frac{a}{2\delta}\right)
\end{equation}
In above equation taking $n_u^c=1$, we get the critical value of $\delta$ in term of $a$ below which 
the model shows abrupt failure.
\begin{equation}\label{eq8}
\delta_c=\displaystyle\frac{a}{2k}.
\end{equation}
Now for the uniform threshold distribution having its mean at $0.5$ we get $a=0.5-\delta$. We thus get the value of critical width: 
\begin{equation}\label{eq9}
\delta_c=\displaystyle\frac{0.5}{2k+1}.
\end{equation}
This result clearly indicates that the tendency of abrupt failure decreases with increasing order $k$ of heterogeneity. 
The existence of such critical disorder was discussed earlier in ref. \cite{Khang}, \cite{Sornette} and \cite{Hansen}. For no heterogeneity, i.e., 
for a single component bundle $\delta_c$ was found to be $1/6$ in ref. \cite{Subhadeep}. We will discuss this point further 
while dealing with numerical results.

%%%%%%%%%%%%%%%%%%%%%%%%%%%%%%%%%%%%%%%%%%%%%%%%%%%%%%%

\section{Numerical Results}
For better understanding of this heterogeneity, we have studied the model numerically. 
A bundle of $L$ fibers is considered for the simulation, with their strengths chosen from a uniform distribution of half width $\delta$ and mean at 0.5. In the mean field limit, we have mainly studied the abruptness in failure process, response of the model to external stress, relaxation dynamics and the burst size distribution. Also we have studied the effect system size and stress release range in presence of local stress concentration. Numerical results are produced with different system sizes, ranging from $10^4$ to $10^5$ over $10^4$ configurations.

%%%%%%%%%%%%%%%%%%%%%%%%%%%%%

\subsection{Mean field limit}
In the mean field limit the load redistribution will be carried 
out through the global load sharing scheme. The fluctuation during the stress redistribution is ignored. 
In this limit the stress redistribution is homogeneous but the stress increment is still heterogeneous, 
as discussed previously. 

%%%%%%%%%%%%%%%

\subsubsection{Failure abruptness}
\begin{figure}[ht]
\centering
\includegraphics[width=7cm, keepaspectratio]{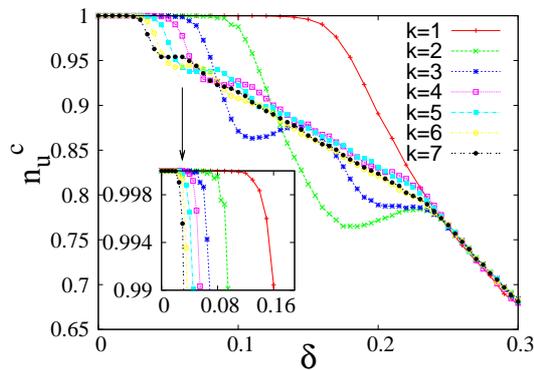}
\caption{Variation of $n_u^c$ with $\delta$, when $k$ different values for local stress are chosen in between 1 and 2. 
The region of $\delta$ where $n_u^c$ starts to deviate from $1$ by an appreciable amount is being zoomed 
and shown in the inset. The zoomed part clearly indicates the decrease in $\delta_c$ with increasing $k$.}
\label{fig:Critical_Point_ManyElasticity}
\end{figure}

The fraction of unbroken fibers just before global failure ($n_u^c$) is studied with a continuous variation of disorder width 
$\delta$ (Fig.\ref {fig:Critical_Point_ManyElasticity}). This study is basically a measurement of abruptness in 
the failure process. $n_u^c=1$ corresponds to brittle like abrupt failure, where no stable state exits.
For $n_u^c<1$, the model undergoes a series of stable states prior to global failure, where an increment of 
applied stress is required to make the model evolve further (like quasi-brittle failure). Previous studies on fiber 
bundle model (in mean field limit) shows that there exists a critical width of disorder $\delta_c$ \cite{Subhadeep}, 
around which the model shows a brittle to quasi-brittle transition. This is the point beyond which $n_u^c$ shows appreciable deviation from $1$ 
and starts decreasing rapidly. Fig.\ref{fig:Critical_Point_ManyElasticity} shows how this $\delta_c$ (the point 
above which $n_u^c$ deviates from 1) shifts with increasing order of heterogeneity $k$. $k=1$ corresponds to the conventional 
fiber bundle model, where stress increment is homogeneous. As $k$ 
increases, $\delta_c$ gradually get shifted to lower and lower value, suggesting decrease in the failure abruptness. 
This means, for higher heterogeneity, the model behaves more like a quasi-brittle material  
rather than the brittle one. In the inset, the part close to $1.0$ is zoomed to show the deviation of $n_u^c$ more clearly.      

%%%%%%%%%%%%%%%

\subsubsection{Response to external stress}
Another way to have a better insight to $\delta_c$ is the response of the model to external stress. At a 
particular external stress per fiber $\sigma$, if the model reaches a stable state after a number of 
redistributing steps, then at that point, the average stress per fiber value (after redistribution) $\langle\sigma_r\rangle$  
will be higher than $\sigma$.
\begin{figure}[ht]
\centering
\includegraphics[width=15cm, keepaspectratio]{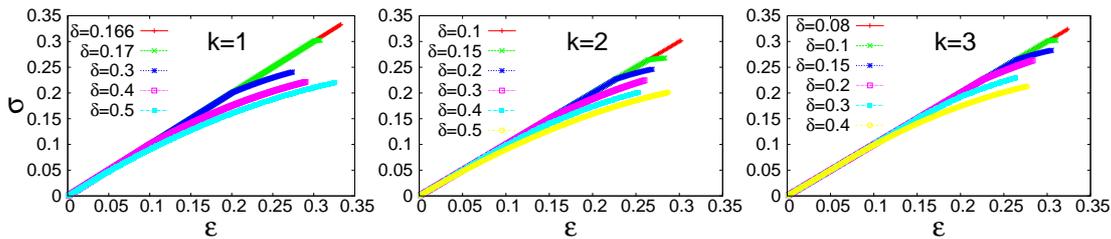}
\caption{Study of stress-strain relationship for the model with $k=1$ (left), 
$k=2$ (center) and $k=3$ (right). Beyond a critical strength of disorder $\delta_c$, a non-linear response is observed.}
\label{fig:Stress_Strain}
\end{figure}
This average stress profile $\langle\sigma_r\rangle$ after redistribution, can be described as the strain 
$\epsilon(\sigma)$ corresponding to the applied stress $\sigma$. A series of such $\sigma$ vs $\epsilon(\sigma)$ values can 
describe the response of the model to external stress. Fig.\ref{fig:Stress_Strain} shows the $\sigma$ 
vs $\epsilon(\sigma)$ behavior for different $k$ values. For $\delta<\delta_c(k)$, the response shows 
purely elastic behavior and there is no deformation observed. Beyond $\delta_c(k)$, the bundle shows 
appreciable non-linear region before global failure. With increasing $k$, the model starts showing this non-linearity 
in relatively lower disorder values. Previous studies on fiber bundle model, with homogeneous stress 
increment ($k=1$), shows this special disorder value ($\delta_c$) to be around $1/6$ for uniform threshold 
distribution \cite{Subhadeep} and $0.23$ for power law distribution \cite{new3}. Fig.\ref{fig:Stress_Strain} shows 
that this $\delta_c$ decreases up to $0.08$ when $k=3$. 

%%%%%%%%%%%%%%%

\subsubsection{Relaxation dynamics}
The existence of $\delta_c$ can also be confirmed from finite size scaling of relaxation time $\tau$ as is done 
in ref. \cite{new1} and ref. \cite{new2}.
\begin{figure}[ht]
\centering
\includegraphics[width=15cm, keepaspectratio]{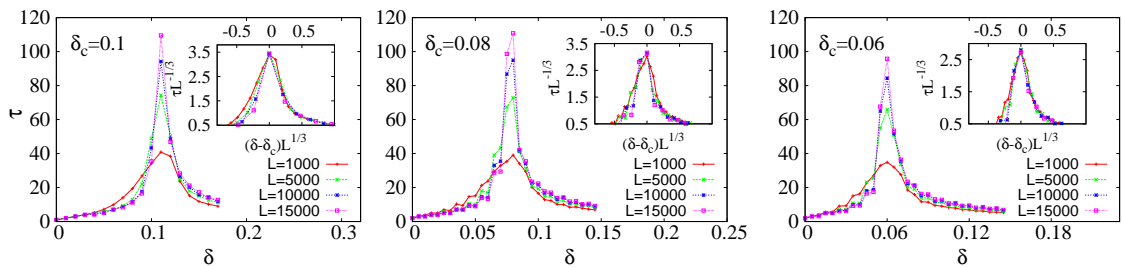}
\caption{Study of $\tau$ with $\delta$ for $k=2$ (left), $k=3$ (center) and $k=4$ (right). 
The system size effect is observed with $L=1000$, $5000$, $10000$ and $15000$. In the inset the figures 
after proper scaling is shown.}
\label{fig:Relaxation_Time}
\end{figure}
For calculating the relaxation time, a minimum stress, sufficient to break the weakest fiber, is applied 
to the model. This rupture of the weakest link might cause further ruptures through redistribution and the model will either break 
completely (if $\delta<\delta_c$) or gradually reach the first stable state (if $\delta>\delta_c$). Number of redistributing steps that the model will go through before its evolution is stopped is defined as the relaxation time ($\tau$) for the model at that $\delta$ value. According to Fig.\ref{fig:Relaxation_Time}, $\tau$ shows a peak at $\delta=\delta_c(k)$ and also diverges with increasing system sizes (same behavior was observed in ref. \cite{Subhadeep} for conventional fiber bundle model). The system size scaling for $\tau$ is shown in the inset. The scaling exponents remain unchanged with change in $k$ values, only the peak shifts to a lower value when $k$ increases.
\begin{equation}
\tau \sim L^{\gamma}\Phi(|\delta-\delta_c(k)|L^{\rho})
\end{equation}
where both $\rho$ and $\gamma$ has the value $1/3$. Only change observer in above behavior is the decrease in 
$\delta_c(k)$ with increasing $k$ values. This shift in $\delta_c$ is consistent with what we observe in the study of stress v/s strain (see Fig.\ref{fig:Stress_Strain}).  

%%%%%%%%%%%%%%%

\subsubsection{Distribution of burst size}
\begin{figure}[ht]
\centering
\includegraphics[width=15cm, keepaspectratio]{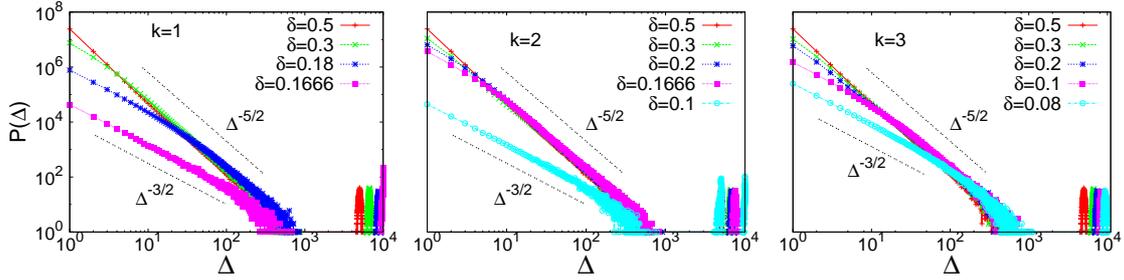}
\caption{Burst size distribution $P(\Delta)$ for $k=1$ (left), $2$ (center), $3$ (right). The distribution shows a scale free behavior where the exponent shows a crossover from value $-5/2$ to $-3/2$ as we approach $\delta_c$ from above. The value of such $\delta_c$ decreases as we increase $k$.}
\label{fig:Burst_Size}
\end{figure}
The other important feature, investigated in our case, is the burst size distribution during the evolution of the bundle. 
A burst is defined as the number of fibers broken in between two stress increment.
In the mean field limit the model is reported to show a 
scale free burst size distribution at $\delta=0.5$ (when the strength values are chosen randomly in between $0$ and $1$) 
with an universal exponent $-5/2$. As the disorder in the model approaches the critical disorder value, the above exponent 
jumps from $-5/2$ to $-3/2$ \cite{new3}. Fig.\ref{fig:Burst_Size} shows such burst size distribution at different order of 
heterogeneity. The crossover from exponent value $-5/2$ to $-3/2$ is observed for $k>1$ also, though this crossover 
takes place at a lower $\delta$ value as we increase $k$.   

%%%%%%%%%%%%%%%

\subsubsection{Probability of abrupt failure}
\begin{figure}[ht]
\centering
\includegraphics[width=7cm, keepaspectratio]{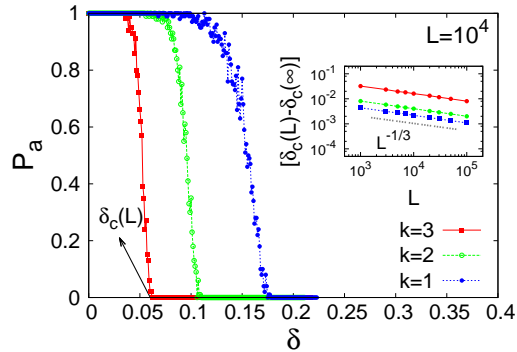}
\caption{Variation of $P_a$, the probability of abrupt failure, with increasing strength of disorder $\delta$. For $\delta>\delta_c(L)$, $P_a=0$ and the failure process is quasi-brittle like non abrupt. $\delta_c(L)$ approaches the thermodynamic limit as: $\delta_c(L)=\delta_c(\infty)+L^{-1/3}$.}
\label{fig:Proability}
\end{figure}

To understand the abruptness in failure process, we have studied the probability $P_a$ of abrupt failure. $P_a$ is basically the ratio of how many times the model breaks abruptly in a single avalanche to the total number of observations. Fig.\ref{fig:Proability} shows the variation of $P_a$ with disorder strength $\delta$. 

At a low $\delta$ value, $P_a=1$ and we observe brittle like abrupt failure at each and every observation. On the other hand, at high disorder the threshold values are are not close to each other and the failure process takes place in a number of avalanches. The brittle to quasi-brittle transition point $\delta_c(L)$ for a certain system size $L$ is defined as the strength of disorder below which there is a non zero probability of abrupt failure. Fig.\ref{fig:Proability} shows that as we increase $k$, $\delta_c(L)$ scales down to lower values and eventually the failure process becomes less abrupt. The inset shows the scaling of $\delta_c(L)$ with increasing system sizes. Specifically, we observe the following scaling
\begin{equation}
\delta_c(L)=\delta_c(\infty)+L^{-\eta}
\end{equation}   
where $\eta=1/3$. $\delta_c(\infty)$ is the brittle to quasi-brittle transition point at the thermodynamic limit. The inset shows that the above scaling of $\delta_c(L)$ remains unchanged even when $k$ is varied.

%%%%%%%%%%%%%%%

\subsubsection{Comparison of analytical and numerical results}
Finally we have reached a point where we can compare the numerical result for $\delta_c(\infty)$ with the analytical expression we obtained for $\delta_c$ as a function of $k$ . 
\begin{figure}[ht]
\centering
\includegraphics[width=6.5cm, keepaspectratio]{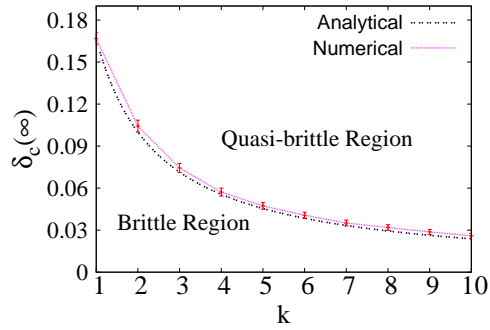}
\caption{Variation of $\delta_c(\infty)$ with $k$ for both numerical (pink) and analytical (black dashed dotted) results.}
\label{fig:Phase_Diagram}
\end{figure}
Fig.\ref{fig:Phase_Diagram} shows the decrease in $\delta_c(\infty)$ with increase in $k$ values. The figure suggests a good agreement between analytical result and the numerical findings. Both suggests that, as we move to a higher order of heterogeneity, the risk of brittle like abrupt failure reduces.

%%%%%%%%%%%%%%%

\subsubsection{High disorder limit}
\begin{figure}[ht]
\centering
\includegraphics[width=7cm, keepaspectratio]{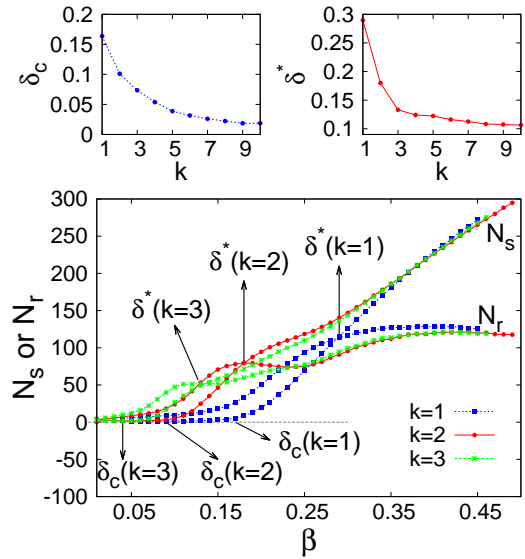}
\caption{Variation of $N_s$ (number of stress increment) and $N_r$ (number of redistribution) 
with disorder width $\delta$. $\delta_c$ separates the brittle region from quasi-brittle. The failure process above 
$\delta^{\ast}$ is mainly guided by stress increment.}
\label{fig:Phase_Diagram1}
\end{figure}

So far we have observed that the model shows brittle like behavior below a certain disorder width, that changes with changing order of heterogeneity. The high disorder limit of this model is still to be explored. Fig.\ref{fig:Phase_Diagram1} shows the variation of $N_s$ (number of stress increment) and $N_r$ (number of redistributing steps), prior to global failure, with increasing $\delta$ values. This interplay of $N_s$ and $N_r$ leads to an upper limit of disorder width $\delta^{\ast}$, beyond which we observe a failure process, mainly guided by external driving force.
For $\delta<\delta_c$, $N_s=1$ and the bundle breaks in a single avalanche (through stress redistribution only), which is a brittle like abrupt failure. In this region, the average avalanche size $\langle s \rangle$ increases linearly with system size $L$ and therefore in the thermodynamic limit there will be an avalanche of infinite size. $\delta_c$ is basically the brittle to quasi-brittle transition point and already discussed in the first part of this paper. For the region $\delta_c<\delta<\delta^{\ast}$, both stress increment and stress redistribution takes place but the failure process is mainly guided by stress redistribution. In this region $\langle s \rangle \sim L^{\zeta}$, with $\zeta$ as a decreasing function of disorder. Finally in the region $\delta>\delta^{\ast}$, stress increment plays the crucial role in the failure process. The $\zeta$ value is even lower here. In a recent paper \cite{mode}, the above behavior has been studied in detail. Here we have observed that as the order of heterogeneity is increased, both $\delta_c$ and $\delta^{\ast}$ get shifted to a lower value. The variation of $\delta_c$ and $\delta^{\ast}$ with order of heterogeneity $k$ is shown in Fig.\ref{fig:Phase_Diagram1}.

%%%%%%%%%%%%%%%%%%%%%%%%%%%%%%%%

\subsection{Local stress concentration}
In this section we have studied a $1$d fiber bundle model with fluctuation in both stress increment and stress redistribution. 
For this purpose we stick to the above mentioned stress increment scheme and we assume that the stress of a broken 
fiber is redistributed uniformly up to $R$ surviving nearest neighbors, known as the stress release range.

%%%%%%%%%%%

\subsubsection{System size effect of critical stress}
A recent study \cite{SroyArxiv} has already described the effect of disorder with local stress concentration. In this paper we will mainly focus on the role of heterogeneity in stress increment, while the disorder width is kept constant by chosing $\delta=0.5$ (the threshold strength values are chosen randomly in between 0 and 1). The results are compared with the results \cite{Chak2,soumya} of conventional fiber bundle model, obtained with above strength of disorder.
\begin{figure}[ht]
\centering
\includegraphics[width=7cm, keepaspectratio]{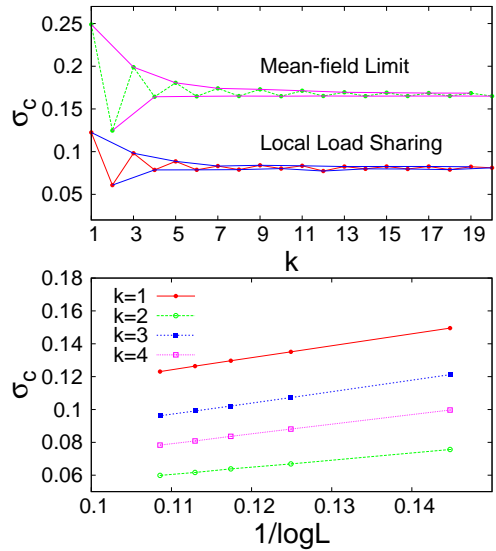}
\caption{Upper: behavior of critical stress $\sigma_c$ with continuous variation of $k$, both in the 
mean-field limit as well as with local stress concentration. $\sigma_c$ decreases with increasing $k$ values. 
Although the $\sigma_c$ for odd $k$ values are relatively higher.
Lower: the behavior of $\sigma_c$ with system size remains unchanged ($\sigma_c \sim 1/\log L$) when $k$ is varied.}
\label{fig:System_Size_Effect}
\end{figure}
We have studied the system size dependence of the critical stress for stress release range $R=1$. The stress release range is basically the number of fibers that carries the stress of the broken fiber. The limit $R=1$ coincides with the LLS limit (shown in Fig.\ref{fig:fbmThres}) as the stress is redistributed between the first surviving nearest neighbor on either side of the broken fiber. The reason behind choosing $R=1$ is, it is the most localized situation and the system size effect is most evident here. With increasing $R$, the model approaches the mean-field limit \cite{soumya} and the system size effect gradually vanishes. The strength of the bundle is observed to decrease in this limit as follows
\begin{equation}
\sigma_c \sim 1/\log L
\end{equation}
Above equation suggests that at thermodynamic limit the bundle will break even at zero stress. The behavior remains 
unaltered when we change the $k$ value, though the exact strength value is observed to alter as we increase $k$ 
(see Fig.\ref{fig:System_Size_Effect}). Fig.\ref{fig:System_Size_Effect} also suggests that, the strength for odd $k$ 
values are obtained to be higher than the even $k$ scenario. $\sigma_c$ oscillates around a particular value, with an 
amplitude gradually decreasing with $k$. This different behavior of $\sigma_c$ for even and odd $k$ is not 
being understood fully and requires further observation. 

%%%%%%%%%%%

\subsubsection{Correlation among rupturing events}
A recent work with variable stress release range shows that there exists a length scale $R_c$ that separates the correlated nucleating failure from the uncorrelated random rupture events. 
\begin{figure}[ht]
\centering
\includegraphics[width=7cm, keepaspectratio]{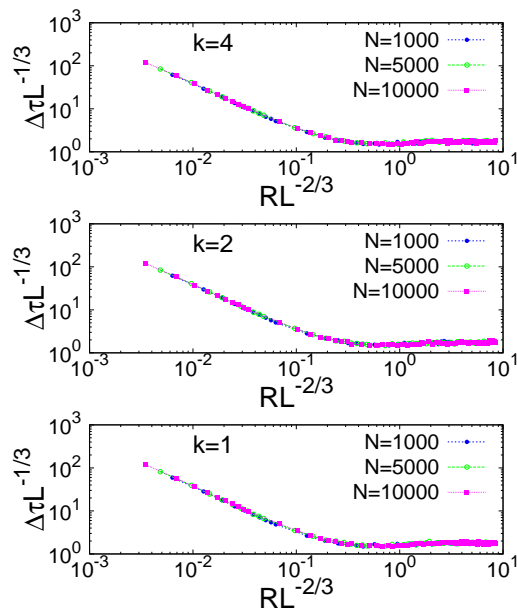}
\caption{Scaling of $\Delta\tau$, time required for final nucleation, with system size $L$ and stress 
release range $R$. We obtain the scaling: $\Delta\tau \sim L^{1/3}\Phi(R/L^{2/3})$, independent of the $k$ value.}
\label{fig:Critical_Range}
\end{figure}
This $R_c$ scales with system size as $L^{2/3}$ \cite{soumya}. For $R>L^{2/3}$, the rupture events are spatially uncorrelated. On the other hand for $R<L^{2/3}$, the crack propagates in nucleating pattern from a single point. The scaling was obtained by observing the time for final nucleation $\Delta\tau$ (number of redistributing steps in between the last stress increment and global failure) with varying $L$ and $R$. 
\begin{equation}
\Delta\tau \sim L^{1/3}\Phi(R/L^{2/3})
\end{equation}
The scaling finally gives us: $R_c \sim L^{2/3}$. We observe that, the scaling remains invariant with respect to the 
order $k$ (see Fig.\ref{fig:Critical_Range}).
$k=1$ corresponds to the conventional fiber bundle model. For $k>1$, there is a fluctuation in the local stress 
profile whenever the stress is increased. The amount of such fluctuation changes when we go to higher $k$ values. 

With local load sharing scheme we have not observed much changes in the conventional results ($\sigma_c\sim 1/\log L$ or $R_c \sim L^{2/3}$). Since in case of LLS scheme the stress redistribution is heterogeneous, a fluctuation in local stress profile is already present in the model. An increasing $k$ value only adds to the pre-existing fluctuation. We will be discussing this LLS scheme in the next section where the continuous limit of the heterogeneous loading is explored.  

%%%%%%%%%%%%%%%%%%%%%%%%%%%%%%%%%

\subsection{The continuous limit}
Finally we have studied the continuous limit of this heterogeneous loading. To construct the continuous 
limit, we consider that each fiber comes with an individual amplification factor $\alpha(i)$, chosen randomly from an uniform distribution with minimum at $1$ and width $\xi$. The local stress profile then follows same Eq.\ref{eq0} and determined by $\sigma$ and different $\alpha(i)$ values associated to each fiber.  
\begin{figure}[ht]
\centering
\includegraphics[width=7cm, keepaspectratio]{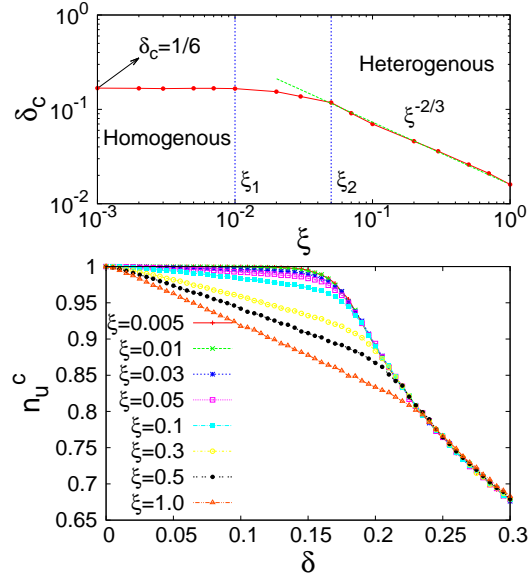}
\caption{$n_u^c$ vs $\delta$ behavior with $\xi$ ranging in between 0.005 and 1.0. $\delta_c$ saturates 
at value $1/6$ for $\xi<0.01$ (homogeneous region). For $\xi>0.5$ (heterogeneous region), $\delta_c$ decreases in a scale free manner. }
\label{fig:Deltac_Continuous}
\end{figure}
The $\alpha(i)$ values lie between 1 and $1+\xi$. For $\xi=0$, the model reaches the conventional limit. Fig.\ref{fig:Deltac_Continuous} shows a continuous variation of $\delta_c$ with $\xi$. The results for this continuous 
limit can be summarized as follows: 
\begin{itemize}
\item For $\xi=0$, $n_u^c$ deviates from 1 at $\delta=\delta_c=1/6$.
\item For $\xi<\xi_1$, the model hardly shifts from the $\xi=0$ limit. The $\delta_c$ value throughout this 
region is almost remains at $1/6$. In this region, the behavior of the bundle does not reflect any heterogeneity.   
\item In the region $\xi_1<\xi<\xi_2$, $\delta_c$ deviates from $1/6$ very slowly.
\item Beyond $\xi_2$, $\delta_c$ decreases in a scale free manner with $\xi$: $\delta_c \sim \xi^{-2/3}$. 
Thus for $\xi>\xi_2$ the model clearly shows the effect of heterogeneity in stress increment.    
\end{itemize}
\begin{figure}[ht]
\centering
\includegraphics[width=7cm, keepaspectratio]{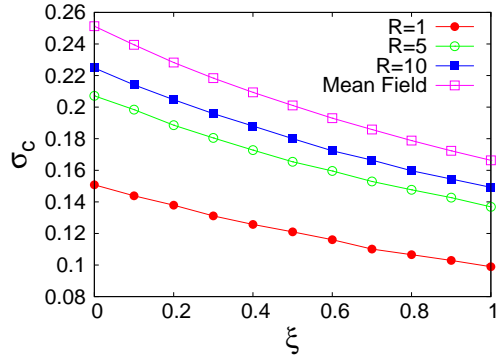}
\caption{Variation of $\sigma_c$ with $\xi$. The results can be compared with Fig.\ref{fig:System_Size_Effect} where the amplification factors are discrete. The system size is kept constant at $L=10^5$.}
\label{fig:Continuous_LLS}
\end{figure}

Fig.\ref{fig:Continuous_LLS} shows the behavior critical stress in local load sharing scheme when the amplification factors $\alpha(i)$'s are continuously distributed over a width $\xi$. The system size is kept constant at $L=10^5$. We have already seen that when the amplification factors have k discrete functions $\sigma_c$ shows a zig zag behavior rather than a monotonic change. Such zig zag behavior vanishes as we enter the continuous limit of such amplification. Above figure shows $\sigma_c$ v/s $\xi$ for $R=1$, $5$, $10$ and $R>R_c$ (the mean field limit). $\sigma_c$ gradually decreases as we increase the width $\xi$. Also with increasing $R$ the model approaches the mean field limit where $\sigma_c=1/4$ for $\xi=0$ and the system size effect vanishes as well. For low $R$, the system size effect of $\sigma_c$ remains unchanged: $\sigma_c \sim 1/\log L$; even when the amplification factors are continuously distributed. 

%%%%%%%%%%%%%%%%%%%%%%%%%%%%%%%%%%%%%%%%%%%%%%%%%%%%%%%%%%%%%%%%%%%%%%%%%%%%%%%%%

\section{Universality}
To check the universal behavior of our results, we have considered a scale free distribution to assign thresholds to individual fibers. Specifically the distribution is given by $P(\sigma)=\sigma^{-1}$ within the window [$10^{\beta}$,$10^{-\beta}$], where $\beta$ determines the amount of disorder. Our findings remain unchanged irrespective of the choice of threshold distribution. \\

\begin{figure}[ht]
\centering
\includegraphics[width=7cm, keepaspectratio]{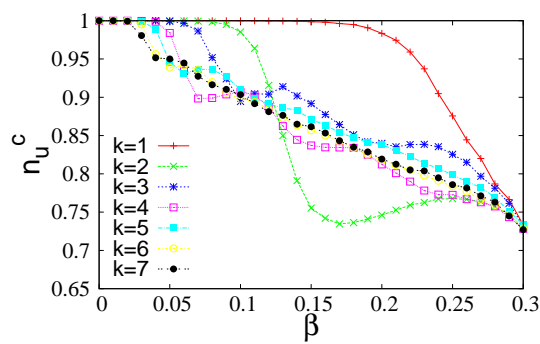}
\caption{Variation of $n_u^c$ with $\beta$ for increasing values of $k$. The $\beta_c$ value decreases as we increase the order of heterogeneity in loading process.}
\label{fig:Critical_Point_ManyElasticity_Powerlaw}
\end{figure}

Fig.\ref{fig:Critical_Point_ManyElasticity_Powerlaw} shows the variation of critical fraction unbroken $n_u^c$ against the disorder $\beta$. The notion of critical disorder is same as previous. $\beta_c$ is the disorder beyond which $n_u^c$ deviates from 1. Above figure clearly shows that $\beta_c$ scales down to a relatively lower value as we increase $k$. This behavior is same as it was in case of uniform distribution. \\

\begin{figure}[ht]
\centering
\includegraphics[width=7cm, keepaspectratio]{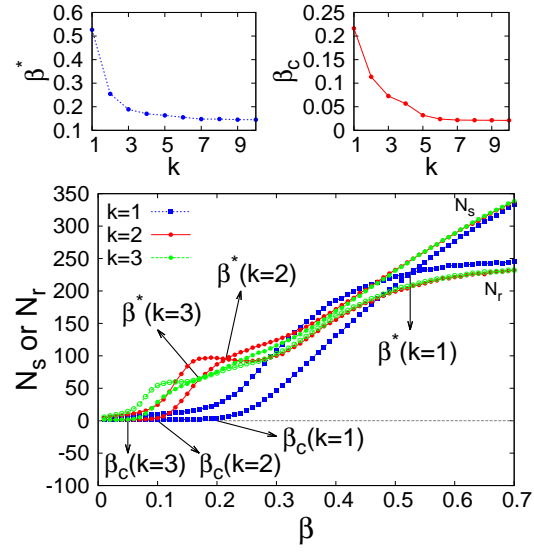}
\caption{Variation of $N_s$ (number of stress increment) and $N_r$ (number of redistribution) 
with disorder $\beta$. $\beta_c$ divides the brittle region from quasi-brittle. The failure process above 
$\beta^{\ast}$ is mainly guided by stress increment. }
\label{fig:Phase_Diagram1_Powerlaw}
\end{figure}

Along with $\beta_c$ we have also checked how $\beta^{\ast}$ (similar to $\delta^{\ast}$ in Fig.\ref{fig:Phase_Diagram1}) changes as we increase the order of heterogeneity. Similar to uniform distribution, in this case also $\beta^{\ast}$ is a decreasing function of $k$. This further suggests less brittle and quasi-brittle response and more temporally uncorrelated events as we increase $k$. Variation of $\beta_c$ and $\beta^{\ast}$ for power law distribution is shown in figure \ref{fig:Phase_Diagram1_Powerlaw}. \\

\begin{figure}[ht]
\centering
\includegraphics[width=7cm, keepaspectratio]{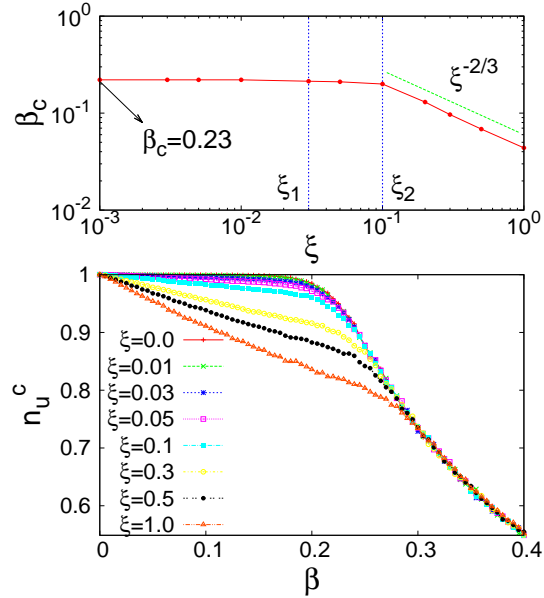}
\caption{The continuous limit of the heterogeneous loading. With increasing $\xi$, the model goes from homogeneous to heterogeneous limit crossing two particular values $\xi_1$ and $\xi_2$.}
\label{fig:Betac_Continuous}
\end{figure}

The continuous limit of the model shows similar results with both the distributions. There is a homogeneous region for $\xi<\xi_1$ and a heterogeneous region beyond $\xi_2$. For $\xi>\xi_2$, $\beta_c$ falls with $\xi$ in a scale-free manner. The exponent of such scale free decrease also shows an universal behavior (See figure \ref{fig:Betac_Continuous}).  
 
Apart from the scale-free distribution, the universality of the results are also confirmed from a truncated Gaussian and truncated Weibull distribution. The strength of disorder for above two distributions are measured from the variance (in case of Gaussian) and Weibull parameter respectively.   

%%%%%%%%%%%%%%%%%%%%%%%%%%%%%%%%%%%%%%%%%%%%%%%%%%%%%%%%%%%%%%%%%%%%%%%%%%%%%%%%%

\section{Discussion}
In this work, the effect of heterogeneous stress increment and heterogeneous stress redistribution is explored in fiber bundle model. In the mean field limit, the order $k$ of heterogeneous loading affects the failure abruptness and changes the brittle to quasi-brittle transition point. The transition point is confirmed from divergence of relaxation time, the failure abruptness and the avalanche size distribution. With local stress concentration, we hardly observe any role of this heterogeneity on system size effect of strength or on the spatial correlation in rupturing process. 
Finally, in the continuous limit of this stress increment scheme, the homogeneous region is observed to be separated from the heterogeneous one around a particular length scale. Such length scale can be expressed in terms of the width $\xi$ of dispersion in amplification factor. Our findings are universal with respect to the choice of the distribution to assign threshold to an individual fiber.

%%%%%%%%%%%%%%%%%%%%%%%%%%%%%%%%%%%%%%%%%%%%%%%%%%%%%%%%%%%%%%%%%%%%%%%%%%%%%%%%%

\section{\textbf{Acknowledgement}} The authors thank Purusattam Ray and Soumyajyoti Biswas for some delightful 
comments and discussions. SR acknowledges Earthquake Research Institute, University of Tokyo for funding during the work.  

%%%%%%%%%%%%%%%%%%%%%%%%%%%%%%%%%%%%%%%%%%%%%%%%%%%%%%%%%%%%%%%%%%%%%%%%%%%%%%%%%

%%%%%%%%%%%%%%%%%%%%%%%%%%%%%%%%%%%%%%%%%%%%%%%%%%%%%%%%%%%%%%%%%%%%%%%%%%%%%%%%%


\begin{thebibliography}{}
\bibitem{Wonga} T. Wonga and P. Baudb, Journal of Structural Geology, vol {\bf 44}, 25-53 (2012). \\
\bibitem{Brede} M. Brede and P. Haasen, Acta Metallurgica, Volume {\bf 36}, Issue 8, pp. 2003-2018 (1988). \\
\bibitem{Gumbsch} P. Gumbsch, J. Riedle, A. Hartmaier, H. F. Fischmeister, Science, Vol. {\bf 282}, Issue 5392, pp. 1293-1295 (1998). \\
\bibitem{Li} R. Li, K. Sieradzki, Phys. Rev. Lett. {\bf 68}, 1168 (1992). \\ 
\bibitem{Khantha1} M. Khantha, D.P. Pope, V. Vitek, Materials Science and Engineering A {\bf 234-236}, 629-632 (1997). \\
\bibitem{Khantha2} M. Khantha, D.P. Pope, V. Vitek, Scripta Metallurgica et Materialia, Vol. {\bf 31}, No. 10, pp. 1349-1354 (1994). \\
\bibitem{Curtin1} W. A. Curtin, Phys. Rev. Lett. {\bf 80}, 1445 (1998). \\
\bibitem{Curtin2} W. A. Curtin and H. Scher, Phys. Rev. Lett. {\bf 67}, 2457 (1993). \\
\bibitem{Nukula} P. K. V. V. Nukula, S. imunovi, S. Zapperi, J. Stat. Mech {\bf P08001} (2004). \\
\bibitem{Khang} B. Kahng, G. G. Batrouni, S. Redner, L. de Arcangelis, H. J. Herrmann, Phys. Rev. B {\bf 37}, 7625 (1988). \\
\bibitem{Shekhawat} A. Shekhawat, S. Zapperi, J. P. Sethna, Phys. Rev. Lett. {\bf 110}, 185505 (2013). \\
\bibitem{Moreira} A. A. Moreira, C. L. N. Oliveira, A. Hansen, N. A. M. Araujo, H. J. Herrmann, J. S. Andrade, Jr., Phys. Rev. Lett. {\bf 109}, 255701 (2012). \\
\bibitem{Chak2} S. Pradhan, A. Hansen and B. K. Chakrabarti, Rev. Mod. Phys. {\bf 82}, 499 (2010). \\
\bibitem{Fiber1} A. Hansen, P. C. Hemmer and S. Pradhan, {\it The Fiber Bundle Model: Modeling Failure in Materials}, WILEY-VCH (2015). \\
\bibitem{Fiber2} S. Biswas, P. Ray and B. K. Chakrabarti, {\it Statistical Physics of Fracture, Beakdown, and Earthquake: Effects of Disorder and Heterogeneity}, WILEY-VCH (2015). \\
\bibitem{Pierce} F. T. Pierce, J. Text. Ind. {\bf 17}, 355 (1926). \\
\bibitem{Tanaka} M.Tanaka, R.Kato, and A.Kayama, Journal of materials science \textbf{37}, 3945-3951 (2002). \\
\bibitem{McClintock} F. A. McClintock, \textit{Statistics of Brittle Fracture, in The Fracture Mechanics of Ceramics}, edited by Bradt R. C. et al., Vol. I (Springer-Verlag, US), p. 93 (1974). \\
\bibitem{Ray} P. Ray and B. K. Chakrabarti, Solid State Commun., {\bf 53}, 477 (1985). \\
\bibitem{Hemmer} P. C. Hemmer and A. Hansen, J. Appl. Mech. {\bf 59}, 909 (1992). \\
\bibitem{lls1} Gomez et al. Phys. Rev. Lett. {\bf 71}, 380 (1993). \\
\bibitem{lls2} Pradhan and Chakrabarti, Int. J. Mod. Phys. B {\bf 17}, 5565 (2003). \\
\bibitem{Daniels} H. E. Daniels, Proc. R. Soc. London, Ser. A {\bf 183}, 405 (1945). \\
\bibitem{Phoenix1} S. L. Phoenix, Adv. Appl. Probab. {\bf 11}, 153 (1979). \\
\bibitem{Phoenix2} R. L. Smith and S. L. Phoenix, J. Appl. Mech. {\bf 48}, 75 (1981). \\ 
\bibitem{Phoenix3} W. I. Newman and S. L. Phoenix, Phys. Rev. E {\bf 63}, 021507 (2001). \\
\bibitem{phoenix09} S.L. Phoenix and W.I. Newman, Phys. Rev. E. {\bf 80}, 066115 (2009). \\
\bibitem{Harlow1} D. G. Harlow and S. L. Phoenix, J. Compos. Mater. {\bf 12}, 314 (1978). \\ 
\bibitem{Harlow2} D. G. Harlow and S. L. Phoenix, Adv. Appl. probab. {\bf 14}, 68 (1982). \\
\bibitem{Harlow3} R. L. Smith, Proc. R. Soc. London, Ser. A {\bf 382}, 179 (1982). \\
\bibitem{soumya} S. Biswas, S. Roy and P. Ray, Phys. Rev. E {\bf 91}, 050105(R) (2015). \\
\bibitem{Raju} I. S. Raju and J. C. Newman Jr., Eng. Frac. Mech. Vol {\bf 11}, 817-829 (1979). \\
\bibitem{Newman} J. C. Newman Jr. and I. S. Raju, Eng. Frac. Mech. Vol {\bf 15}, 185-192 (1981). \\ 
\bibitem{Vainshtok} V. A. Vainshtok and I. V. Varfolomeyeb, Eng. Frac. Mech. Vol {\bf 34}, 125-136 (1989). \\
\bibitem{Wang} X. Wang and S. B. Lambert, Eng. Frac. Mech. Vol {\bf 51}, No. 4, 517-532 (1995). \\
\bibitem{Fett} T. Fett, Eng. Frac. Mech. Vol {\bf 36}, No. 4, 647-651 (1990). \\
\bibitem{phoenix73} S. L. Phoenix and H. M. Taylor, Adv. Appl. Prob. {\bf 5}, 200-216 (1973). \\
\bibitem{phoenix75} S. L. Phoenix, Int. J. Engng. Sci. Vol {\bf 13}, 287-304 (1975). \\
\bibitem{phoenix74} S. L. Phoenix, Fiber Science and Technology Vol {\bf 7}, 15-31 (1974). \\
\bibitem{daniels89} H.E. Daniels, Adv. Appl. Prob. {\bf 21}, 315-333 (1989). \\
\bibitem{phoenix79} S. L. Phoenix, {\it Statistical aspects of failure of fibrous materials}, Composite Materials: Testing and Design (Fifth Conference, {\bf STP674}) edited by S. W. Tsai (1979). \\
\bibitem{Sornette} J. V. Andersen,  D. Sornette and K. T. Leung, Phys. Rev. Lett. {\bf 78}, 2140 (1997). \\
\bibitem{Hansen} S. Pradhan and A. Hansen, Phys. Rev. E {\bf 72}, 026111 (2005). \\
\bibitem{Subhadeep} S. Roy and P. Ray, Europhys. Lett. {\bf 112}, 26004 (2015). \\
\bibitem{new3} C. Roy, S. Kundu, and S. S. Manna, Phys. Rev. E {\bf 91}, 032103 (2015). \\
\bibitem{new1} S. Pradhan, P. Bhattacharyya, and B. K. Chakrabarti, Phys. Rev. E {\bf 66}, 016116 (2002). \\ 
\bibitem{new2} C. Roy, S. Kundu, and S. S. Manna, Phys. Rev. E {\bf 87}, 062137 (2013). \\
\bibitem{mode} S. Roy, S. Biswas and P. Ray, arXiv:{\bf 1610.06942v2} (2017). \\
\bibitem{SroyArxiv} S. Roy, Phys. Rev. E \textbf{96}, 042142 (2017). 
\end{thebibliography}
\end{document}